# On the importance of the electron-phonon coupling function on the superconducting transition temperature in dodecaboride superconductors: A comparison of $LuB_{12}$ with $ZrB_{12}$.

J. Teyssier[‡], R. Lortz, A. Petrovic, D. van der Marel

*Département de Physique de la Matière Condensée, Université de Genève, Quai Ernest-Ansermet 24, 1211 Genève 4, Switzerland*

V. Filippov, N. Shitsevalova

*Institute for Problems of Materials Science, National Academy of Sciences of Ukraine, 252680 Kiev, Ukraine*

We report a detailed study of specific heat, electrical resistivity and optical spectroscopy in the superconducting boride $LuB_{12}$ ($T_c$ = 0.4 K) and compare it to the higher $T_c$ compound $ZrB_{12}$ ($T_c$ = 6 K). Both compounds have the same structure based on enclosed metallic Lu or Zr ions in oversized boron cages. The infrared reflectivity and ellipsometry in the visible range allow us to extract the optical conductivity from 6 meV to 4 eV in the normal state from 20 to 280 K. By extracting the superconducting properties, phonon density of states and electron-phonon coupling function from these measurements we discuss the important factors governing $T_c$ and explain the difference between the two compounds. The phonon density of states seems to be insignificantly modified by substitution of Zr with Lu. However, the soft vibrations of the metal ions in boron cages, responsible for the relatively high $T_c$ in $ZrB_{12}$, have almost no contribution to the electron-phonon coupling in $LuB_{12}$.

**Keywords:** $LuB_{12}$, $ZrB_{12}$, borides, superconductivity, specific heat, resistivity, phonons, optical spectroscopy

**PACS numbers:** 74.70.Ad, 74.25.Bt, 72.15.Eb, 78.20.Ci, 78.30.-j

## I. Introduction

The discovery of superconductivity in $MgB_2$ has stimulated intense research on other superconducting borides. The superconducting borides with the second and third highest transition temperatures are $YB_6$ ($T_c$ = 6 - 7.5 K) )[1] and $ZrB_{12}$ ($T_c$ = 6 K), respectively[2]. $T_c$'s of more than one order of magnitude lower are found in $LaB_6$ ($T_c$ < 0.1 K)[3] and $LuB_{12}$ ($T_c$ = 0.44 K)[4,5]. The two latter compounds belong to the rare earth boride family, intensively studied because of their wide variety of physical properties[6]: valence fluctuations ($SmB_6$, $YbB_{12}$), magnetic ordering ($CeB_6$, $HoB_{12}$) and superconductivity ($LaB_6$, $LuB_{12}$). ). Rare earth hexa- and dodecaborides both exhibit structures based on oversized crystalline cages formed by a covalent 2p boron ions in which metal atoms are enclosed. Such materials are particularly interesting when they become superconducting, since they then represent model systems for studying electron-phonon interactions[7,8]. Apart from the borides, superconductors with similar structures are found for example among the pyrochlores[9,10] or the clathrates[11-14]. Strong covalent bonds between the boron atoms in $B_{12}$ cages lead to a very rigid lattice with a high Debye temperature. The size of the caged atom only has a minor effect on the lattice parameter; however the presence of the metal atoms stabilizes the structure since two valence electrons per metal atom are transferred to the cages in order to compensate the electron deficiency in the boron sublattice The enclosed ions undergo large vibrations in form of soft Einstein phonons which have been reported to mediate superconductivity in $YB_6$[8] and $ZrB_{12}$[7].

---

[‡] Corresponding author: e-mail: Jeremie.Teyssier@physics.unige.ch

Since large high-quality single crystals are available, the superconductivity in $ZrB_{12}$ and $YB_6$ has been object of intense research. The isotope effect in $ZrB_{12}$ for zirconium[15] ($\beta \approx -0.32$) is much larger than for boron[16] ($\beta \approx -0.09$), pointing to a large contribution of lattice modes involving Zr atoms to the electron-phonon coupling responsible for superconductivity in $ZrB_{12}$. The same was concluded from an inversion of specific heat and resistivity data for $YB_6$[8] and $ZrB_{12}$[7] and the coupling was mainly attributed to these low energy phonon modes. An absence of magnetism in $LuB_{12}$ lead to its use as a non-magnetic reference for the study of the Kondo insulator $YbB_{12}$[17]. $LuB_{12}$ and $ZrB_{12}$ band structure calculations[18-20] revealed very similar total densities of states at the Fermi level. The lattice dynamic were also studied showing closely-related phonon spectra for both compounds[21]. Considering these similarities in the crystal dynamics and electronic structures, the large difference in the superconducting transition temperature in $LaB_6$, $YB_6$ and $ZrB_{12}$, $LuB_{12}$ pairs appears incomprehensible. The hexaborides have been studied by Schell et al.[22] but up until now the dodecaborides have not yet been investigated.

In this paper, we present specific heat and electrical resistivity experiments used as "thermal spectroscopies" in combination with optical measurements on $LuB_{12}$. We discuss the electron-phonon coupling strength in this material and compare it to $ZrB_{12}$ for which we recently reported thermodynamic[7] and optical[20] investigations of the Eliashberg function. Our experiments reveal that the electron-phonon coupling of the vibration of the Lu ions in the boron cages is almost absent, in direct contrast with the particularly strong coupling in the higher $T_c$ compound $ZrB_{12}$.

## II. Techniques

$LuB_{12}$ crystallizes in the $UB_{12}$ type structure work[23] which can be viewed as a cubic rocksalt arrangement of Lu and $B_{12}$ cuboctahedral clusters. The sample preparation process involves synthesising dodecaborides by a borothermal reduction of the metal oxides in vacuum at 1900 K, compacting these powders into rods and subsequently sintering them, then finally inductive zone melting in a high frequency induction zone unit[24].

The specific heat was measured using a high-precision continuous-heating adiabatic calorimeter between 16 and 300 K and a relaxation calorimeter in a $^3$He cryostat in the temperature range between 300 mK and 16 K. The resistivity was measured using a standard four probe technique.

Optical measurements were performed on the (001) surface of a single crystal of diameter 5 mm. In the visible photon energy range (0.8 - 4 eV), the complex dielectric function $\varepsilon(\omega)$ was determined directly using spectroscopic ellipsometry at an incident angle of 62°. The reflectivity of the sample was measured in the infrared spectral range (6 meV and 0.8 eV) using a Bruker 113 Fourier transform spectrometer. The reflectivity reference was taken by *in situ* gold evaporation. For the optical experiments the sample was mounted in a helium flow cryostat allowing measurements from room temperature down to 10 K. Fig. 1a presents the real and imaginary parts of $\varepsilon(\omega)$ measured by ellipsometry at selected temperatures. The reflectivity measured at low frequencies and extracted from the dielectric constant in the visible range, are plotted in Fig. 1b.

A mismatch of 1.4% is observed between the reflectivity extracted from ellipsometry data in the visible range and the reflectivity measured in the infrared. This mismatch has not been corrected and is revealed as a jump in the optical conductivity (Fig. 2). In order to obtain the optical conductivity in the infrared region we used a variational routine[26] yielding the Kramers-Kronig consistent dielectric function which reproduces all the fine details of the infrared reflectivity data while *simultaneously* fitting to the complex dielectric function in the visible and UV-range.

Fig. 2 shows the evolution of the optical conductivity with temperature. A Drude-like peak, which narrows when the temperature is reduced, indicates metallic behavior. The DC conductivity is also displayed in this graph. LuB$_{12}$ is a very good conductor with a resistivity at 20 K of $\rho_{DC}$ = 2.05 $\mu\Omega$ cm.

## III. Superconducting properties

The specific-heat of LuB$_{12}$ has been previously reported[5,27]. Fig. 3 presents the specific heat $C/T$ of LuB$_{12}$ in comparison to that of ZrB$_{12}$ reported in Ref.[7] in the temperature range from 350 mK to 10 K on a logarithmic temperature scale. Sharp jumps indicate the superconducting transition temperatures at 6 K (ZrB$_{12}$) and 0.42 K (LuB$_{12}$). The upturn below 2 K above $T_c$ in LuB$_{12}$ is related to a Schottky anomaly which we attribute to nuclear moments with a characteristic energy of 0.9 K (indicated by the dotted line).

A magnetic field of 1 T is sufficient to suppress superconductivity. This allows us to analyze the normal-state specific heat in a standard way according to the expansion:

$$C_n(T \to 0) = \gamma_n T + \sum_{n=1}^{3} \beta_{2n+1} T^{2n+1},$$

where the first term is the electronic contribution, with $\gamma_n = \frac{1}{3}\pi^2 k_B^2 (1+\lambda_{ep}) N(E_F)$, $k_B$ Boltzmann's constant, $\lambda_{ep}$ the electron-phonon coupling constant and $N(E_F)$ the band-structure density of states at the Fermi level including two spin directions [i.e. the electronic density of states (EDOS)]. The second term is the low-temperature expansion of the lattice specific heat, where $\beta_3 = \frac{12}{5} N_{Av} k_B \pi^4 \theta_D^{-3}(0)$, with $N_{Av}$ Avogadro's number and $\theta_D(0)$ the initial Debye temperature. From a fit from 2.5 K to 5 K we obtain $\gamma_n$ = 0.26 mJ gat$^{-1}$K$^{-2}$. This value is slightly smaller than values reported in literature[5,27]. The Sommerfeld constant corresponds to a dressed density of states at the Fermi level $(1+\lambda_{ep})N(E_F)$ = 0.11 states eV$^{-1}$ atom$^{-1}$ which is slightly lower than the value of ZrB$_{12}$. A comparison with recent band structure calculations[28] (see Table I) leaves space for a small renormalization factor of $\lambda_{ep} \cong 0.41$ (LuB$_{12}$) and $\lambda_{ep} \cong 0.38$ (ZrB$_{12}$). Surprisingly, both values are very close to each other. We will discuss this at a later point. Both values are in the weak-coupling regime of superconductivity. The normalized specific-heat jump is $\Delta C/\gamma_n T_c$ = 1.14, which is also in the weak coupling limit below the BCS value of 1.43 and lower than the value reported for ZrB$_{12}$ ($\Delta C/\gamma_n T_c$ = 1.66[7]) thus pointing towards a smaller coupling constant $\lambda_{ep}$. Table I gives an overview of the superconducting properties.

## IV. Phonon density of states and electron phonon coupling

It has been shown previously[7,8,29] that specific heat can provide information usually taken from inelastic neutron scattering with limited but sufficient accuracy to characterize the phonon density of states (PDOS). In a similar manner, an accurate measurement of the resistivity can provide information on the electron-phonon coupling function usually measured by tunneling experiments[7,8,29]. The normal state specific heat of both compounds has been previously investigated[7,8,27], whereas the resistivity of LuB$_{12}$ has not yet been analyzed in this way. In the following we will directly compare the PDOS of both compounds to study how the enclosed ion in the boron cages influences the PDOS and electron-phonon coupling function in order to understand the difference in $T_c$ between the two compounds. The PDOS obtained from the specific heat is needed as an input to extract the electron-phonon coupling function for transport which is closely related to the electron-phonon coupling function for superconductivity.

It has been reported previously that the specific heat in the normal state of both compounds shows a rather unusual temperature dependence at low temperatures (Fig. 4)[7,27]. The low-temperature $T^3$ regime of the lattice specific heat does not extend beyond a few Kelvin, as shown by the large positive curvature of the normal-state curve in Fig. 3. A simplified method of obtaining the PDOS from specific heat consists of representing $F(\omega)$ by a basis of Einstein modes with constant spacing on a logarithmic frequency axis:

$$F(\omega) = \sum_k F_k \delta(\omega - \omega_k). \tag{1}$$

The corresponding lattice specific heat is given by:

$$C_{ph}(T) = 3 N_{Av} k_B \sum_k F_k \frac{x_k^2 e^{x_k}}{(e^{x_k} - 1)^2} \tag{2}$$

where $x_k = \omega_k / T$. The weights $F_k$ are found by a least-squares fit of the lattice specific heat. The number of modes is chosen to be small enough to ensure the stability of the solution; we used $\omega_{k+1}/\omega_k = 1.75$. Note that we do not try to find the energy of each mode; we rather aim to establish a histogram of the phonon density in predefined frequency bins. The robustness of the fit is demonstrated by the r.m.s deviation: <0.4% above 10 K.

Figure 5 illustrates the decomposition of the lattice specific heat into a set of Einstein functions. We use a plot of $C/T^3$ as in this representation contributions related to Einstein phonons appear as a bell-shaped curve[30]. In the case of ZrB$_{12}$, a low lying Einstein phonon is visible with a characteristic energy of 170 K. This mode is also present in LuB$_{12}$ where the mode is slightly shifted down to 162 K in accordance with Ref.[27]. The PDOS obtained in this way is included in Fig. 7. In analogy with other isostructural borides, it consists of a quasi-Debye background with a high characteristic frequency (~1000 K), as expected in view of the boron mass, superimposed on a low-energy mode at 14 - 15 meV, presumably associated with the oscillations of the Lu / Zr atoms in the boron "cages" present in the structure. The nature of the metallic ion only has a minor influence on the PDOS. Replacing Zr by Lu, the low frequency mode shifts down from 15 to 14 meV and its amplitude increases. This small shift in phonon frequency certainly cannot explain the large difference in $T_c$.

In ZrB$_{12}$ it has been previously demonstrated that the ~15 meV mode contributes strongly to the electron-phonon coupling and mediates superconductivity[7]. The question eventually arises whether this coupling is also present in LuB$_{12}$. To investigate this question we follow the same approach as for ZrB$_{12}$ using resistivity and optical spectroscopy as experimental probes.

We analyse the resistivity (Fig. 6) in a similar way to the specific heat. We start from the generalized Bloch-Grüneisen formula (see e.g. Ref.[31], in particular p. 212 and 219):

$$\rho(T) = \rho(0) + \frac{4\pi m^*}{ne^2} \int_0^{\omega_{max}} \alpha_{tr}^2 F(\omega) \frac{xe^x}{(e^x - 1)^2} d\omega \tag{3}$$

where $x \equiv \omega/T$ and $\alpha_{tr}^2 F(\omega)$ is the electron-phonon "transport coupling function" or transport Eliashberg function. In the restricted Bloch-Grüneisen approach, one would have $\alpha_{tr}^2 F(\omega) \propto \omega^4$, and as a consequence $\rho(T) - \rho(0) \propto T^5$, but deviations from the Debye model, complications with phonon polarizations and Umklapp processes would not justify this simplification beyond the low-temperature continuum limit. Using a decomposition into Einstein modes similar to Eq. (2),

$$\alpha_{tr}^2 F(\omega) = \sum_k \alpha_k^2 F_k \delta(\omega - \omega_k), \tag{4}$$

we obtain the discrete version of Eq. (3):

$$\rho(T) = \rho(0) + \frac{4\pi m}{ne^2} \sum_k \alpha_k^2 F_k \frac{x_k e^{x_k}}{(e^{x_k} - 1)^2} \tag{5}$$

where the fitting parameters are the $\alpha_k^2 F_k$ coefficients. The residual resistivity $\rho(0) = 2.05$ $\mu\Omega$ cm was determined separately. Fig. 6 shows the decomposition of the total resistivity into Einstein components. The electron-phonon transport coupling function $\alpha_{tr}^2 F(\omega)$ is closely related to the isotropic Eliashberg function $\alpha^2 F(\omega)$ which governs superconductivity[32]. In ZrB$_{12}$ the main component arises from modes with energies near 170 K, which results in a pronounced peak in $\alpha_{tr}^2 F(\omega)$ at 170 K (~15 meV) (Fig. 7). Compared to other modes, the ~170 K region in ZrB$_{12}$ is weighted much more heavily in the $\alpha_{tr}^2 F(\omega)$ function than in the PDOS $F(\omega)$. In case of LuB$_{12}$ the low frequency mode which is shifted down to 162 K (~14 meV) shows a peak in $\alpha_{tr}^2 F(\omega)$ which is 4 times smaller than that of ZrB$_{12}$. The electron-phonon coupling parameter relevant for transport $\lambda_{tr} \equiv 2\int \omega^{-1} \alpha_{tr}^2 F(\omega)$ is obtained from $\lambda_{tr} = \sum_k \lambda_{tr,k} = 0.42$ for ZrB$_{12}$ and $\lambda_{tr} = \sum_k \lambda_{tr,k} = 0.29$ for LuB$_{12}$. The prefactor is calculated from the plasma frequency which we extracted from optical experiments (Chapter V). In accordance with the value of $\lambda_{ep}$ determined from the Sommerfeld constant, both values are surprisingly close to each other and cannot explain the large $T_c$ difference. This is due to the larger weightening of some high energy modes in LuB$_{12}$, which partially compensates the smaller weight of the 162 K mode. However, if we only consider the low energy electron-phonon coupling parameters $\lambda_{tr,k}$ of the mode associated with the vibration of the Lu and Zr ions in their cages, we find $\lambda_{tr,k} = 0.2$ for ZrB$_{12}$ and $\lambda_{tr,k} = 0.05$ for LuB$_{12}$. Although the difference in the values does not exactly reflect the difference in $T_c$ of more than one order of magnitude, it indicates that it is the lack of electron-phonon coupling of the mode associated with the vibrations of the Lu ion in the cages which is responsible for the low $T_c$ in this compound. Expressed alternatively, it is the particularly strong electron-phonon coupling of the vibrations of the Zr ions in the boron cages which raises the $T_c$ of ZrB$_{12}$ to 6 K.

In the case of a pronounced electron-phonon interaction we have recently shown[20] that it is possible to extract the electron-phonon coupling function $\alpha_{tr}^2 F(\omega)$ from the optical conductivity. Then the simple Drude model becomes inapplicable to the low frequency region. We therefore adopted the following model for the dielectric function:

$$\varepsilon(\omega) = \varepsilon_\infty - \frac{\Omega_p^2}{\omega[\omega + iM(\omega,T)]} + \sum_j \frac{\Omega_{p,j}^2}{\omega_{0,j}^2 - \omega^2 - i\omega\gamma_j} \quad (6)$$

$\varepsilon_\infty$ represents the contribution of core electrons; the second and the third terms describe free carriers and interband contributions respectively. The latter is taken to be a sum of Lorentzians with adjustable parameters. The frequency-dependent scattering of the free carriers is expressed via the memory function:

$$M(\omega,T) = \gamma_{imp} - 2i \int_0^\infty d\Omega\, \alpha_{tr}^2 F(\Omega) K\left(\frac{\omega}{2\pi T}, \frac{\Omega}{2\pi T}\right) \quad (7)$$

where $\gamma_{imp}$ is the impurity scattering rate and $\alpha_{tr}^2 F(\Omega)$ is the transport electron-phonon coupling function. $K$ is the kernel as described by Dolgov et al.[33].

For ZrB$_{12}$, an electron-phonon coupling constant $\lambda \cong 1$ was extracted in this way in agreement with previous reports[34-36]. Fig. 8 shows a comparison at low temperature of the low frequency part of the reflectivity for the two compounds. The fits, including the contribution from the electron-phonon interaction as described above, are presented by a solid red line. For LuB$_{12}$ a good fit of both the ellipsometry and reflectivity data can be obtained using only the standard Drude-Lorentz model. This indicates a very low contribution of phonons to electron scattering. As a consequence, the direct extraction of the transport electron-phonon coupling function from optical data, as for ZrB$_{12}$, would be to imprecise. To nevertheless incorporate the electron-phonon interaction, we inserted the $\alpha_{tr}^2 F(\Omega)$ obtained from resistivity (Fig.7) into Equation 7. In the case of ZrB$_{12}$, an additional electron-boson interaction was necessary to reproduce the experimental data. Parameters relative to the free carriers in eq. 6 (used in the fit) are given in table II. The plasma frequency calculated using LDA is also presented and a similar deviation of about 15% is observed for both compounds.

As for a normal metal, the plasma frequency is slightly increased upon cooling the system because of the narrowing of the Drude peak. It has been shown that the plasma frequencies of these two dodecaborides follow the opposite trend. Even if these materials are extremely good metals in the normal state, the delocalization of the metal ions within the cages (demonstrated by using X-ray diffraction[21]) reduces the number of charge carriers. This decrease in the plasma frequency has been quantitatively related to the delocalization effect through band structure calculations.

## VII. Discussion and conclusions

In order to understand why the $T_c$ of LuB$_{12}$ ($T_c$ = 0.4 K) is so low compared to that of ZrB$_{12}$ ($T_c$ = 6 K), we have performed a comparative analysis of specific-heat, electrical resistivity and optical data. The Sommerfeld constant points towards a dressed density of states $(1+\lambda_{ep})N(E_F)$ = 0.11 states eV$^{-1}$ atom$^{-1}$, which is slightly lower than the value of 0.144 states eV$^{-1}$ atom$^{-1}$ of ZrB$_{12}$. For the electron-phonon coupling parameter $\lambda_{ep}$ we extract surprisingly similar values of $\lambda_{ep} \cong 0.41$ for LuB$_{12}$ and $\lambda_{ep} \cong 0.38$ for ZrB$_{12}$, both lying in the weak-coupling regime of superconductivity. These values are very close to the $\lambda_{tr}$ values $\lambda_{tr} = \sum_k \lambda_{tr,k} = 0.42$ for ZrB$_{12}$ and $\lambda_{tr} = \sum_k \lambda_{tr,k} = 0.29$ for LuB$_{12}$ obtained from the resistivity. The normalized specific-heat jump $\Delta C/\gamma_n T_c$ (which is another measure of the coupling strength) is however clearly smaller in LuB$_{12}$ compared to ZrB$_{12}$ (see Table I), therefore indicating that a weaker electron-phonon coupling may nevertheless be the main reason for the large difference in $T_c$ between the two compounds. As shown previously[7,27], both compounds show a pronounced peak in the PDOS associated with the vibration of the enclosed metal atoms in the cage-like boron host. In ZrB$_{12}$ this Einstein phonon is centered at 170 K; replacing Zr by Lu shifts the mode down to 162 K. This and the only marginally lower density of states at the Fermi level can not only account for the large change of $T_c$ from 6 K down to 0.4 K. We have previously reported that the Einstein phonon at 170 K in ZrB$_{12}$ shows a strong peak in the transport electron-phonon coupling function $\alpha_{tr}^2 F(\omega)$. $\alpha_{tr}^2 F(\omega)$ is closely related to the isotropic Eliashberg function for superconductivity $\alpha^2 F(\omega)$ and a strong peak in $\alpha_{tr}^2 F(\omega)$ therefore indicates the phonon mode which provides most of the superconducting coupling and determines the $T_c$. In the case of LuB$_{12}$ we find that electron-phonon coupling to the corresponding mode at 162 K is nearly absent: the $\alpha_{tr}^2 F(\omega)$ only shows a tiny peak at this energy.

The smallness of the electron-phonon coupling constant in LuB$_{12}$ is confirmed by the optical spectroscopy which shows that the $\alpha_{tr}^2 F(\omega)$ function derived from thermodynamic and transport experiments is entirely realistic.

The main reason for the lower $T_c$ in LuB$_{12}$ is thus the weaker coupling of the Einstein phonon related to the vibration of the enclosed ion to the conduction electrons. The origin of this weaker electron-phonon interaction remains unclear but may be found in the different "volume filling factors" of the boron cages. This factor, which we define as the ratio between the volumes of the caged ions to the total volume of the boron cages, tunes the hybridization of the Lu/Zr electronic orbitals with those of the boron cage atoms and therefore strongly influences the electron-phonon interaction. The optical experiments furthermore reveal an abnormal temperature dependence of the plasma frequency in LuB$_{12}$. Such an effect has already been observed in ZrB$_{12}$ and was attributed to a delocalisation of the Zr ion in the boron cages[20]. The hybridization of the orbitals of the caged atoms with those of the boron cages would react very sensitively to such a delocalization. This may result in a stronger suppression of charge carriers at the Fermi level at low temperatures in LuB$_{12}$.

## Acknowledgments
R.L. thanks A. Junod for sharing his knowledge about calorimetry and its analysis. This work is supported by the Swiss National Science Foundation through grant 200020-109588 and the National Center of Competence in Research (NCCR) 'Materials with Novel Electronic Properties-MaNEP'.

**Tables:**

|  | **LuB$_{12}$** | **ZrB$_{12}$** |
|---|---|---|
| $T_c$ [K] | 0.42 ± 0.05 | 5.96 ± 0.05 |
| $\gamma_n$ [mJ/mole K$^2$] | 0.26 ± 0.02 | 0.34 ± 0.02 |
| $\Delta C/\gamma_n T_c$ | 1.14 ± 0.1 | 1.66 ± 0.1 |
| $2\Delta_0/k_B T_c$ | 3.2 ± 0.1 | 3.7 ± 0.1 |
| $(1+\lambda_{ep})N(E_F)$ [states eV$^{-1}$ atom$^{-1}$] | 0.110 | 0.144 |
| $N(E_F)$ [states eV$^{-1}$ atom$^{-1}$][28] | 0.078 | 0.104 |
| $\lambda_{ep}$ | 0.41 | 0.38 |

**Table I.** Superconducting parameters of LuB$_{12}$ in comparison to ZrB$_{12}$.

|  | **LuB$_{12}$** | **ZrB$_{12}$** |
|---|---|---|
| $\Omega_p$ [eV] | 4.9 | 5.5 |
| $\Omega_p$ LDA [eV] [28] | 5.7 (+16%) | 6.3 (+14%) |
| $\gamma_{imp}$ [meV] | 31 | 33 |

**Table II.** Fitting parameters of the low frequency component of our optical spectra related to the free carriers term in Eq. 6. A good correlation is observed with the plasma frequency predicted by LDA.

**Figure Captions:**

**Fig. 1** a) Temperature dependence of the reflectivity. The dashed line corresponds to the reflectivity measured by Okamura et al.[25]. b) Temperature dependence of the dielectric function measured using ellipsometry.

**Fig. 2.** Temperature dependence of the real part of the optical conductivity. DC values from resistivity measurements are shown in the left panel.

**FIG. 3.** (Color online) Total specific heat $C/T$ of $LuB_{12}$ (closed symbols) in the superconducting states (zero-field data) and the normal states (superconductivity has been suppressed by a magnetic field of 1 T) in comparison to $ZrB_{12}$ (open symbols)[7] showing the superconducting transitions at 0.42 K and 6 K respectively. The dotted line models a Schottky contribution which we attribute to nuclear moments.

**FIG. 4.** (Color online) Total specific heat divided by temperature for $ZrB_{12}$ and $LuB_{12}$.

**FIG. 5.** (Color online) Lattice specific heat divided by $T^3$ of $ZrB_{12}$ and $LuB_{12}$ showing its decomposition into Einstein terms. Our model uses a set of Einstein functions where characteristic temperatures are equally spaced on a logarithmic scale.

**FIG. 6.** (Color online) Total resistivity divided by temperature in $ZrB_{12}$[7] (upper panel) and $LuB_{12}$ (lower panel) showing the decomposition into Einstein terms and the residual term. The largest Einstein component in $ZrB_{12}$ is centered on $\omega = 170$ K. The corresponding mode in $LuB_{12}$ at $\omega = 162$ only shows a small amplitude.

**FIG. 7.** (Color online) Electron-phonon transport coupling function $\alpha_{tr}^2 F(\omega)$ (closed circles) of $ZrB_{12}$ (a) and $LuB_{12}$ (b) deconvolved from the resistivity in comparison to the phonon density of states $F(\omega)$ deconvolved from the specific heat (histogram of rectangles). Fits are performed with δ-functions $(\alpha_{tr}^2 F)_k \delta(\omega - \omega_k)$ on a basis of Einstein frequencies $\omega_{k+1} = 1.75 \omega_k$.

**Fig. 8.** Low frequency part of the reflectivity data at 25 K for $ZrB_{12}$ and 20K for $LuB_{12}$ (symbols). Dashed blue curves are fits to experimental data with a simple Drude model. Red solid lines are fits taking into account the electron phonon interaction given by the transport electron-phonon coupling function presented in Fig.7.

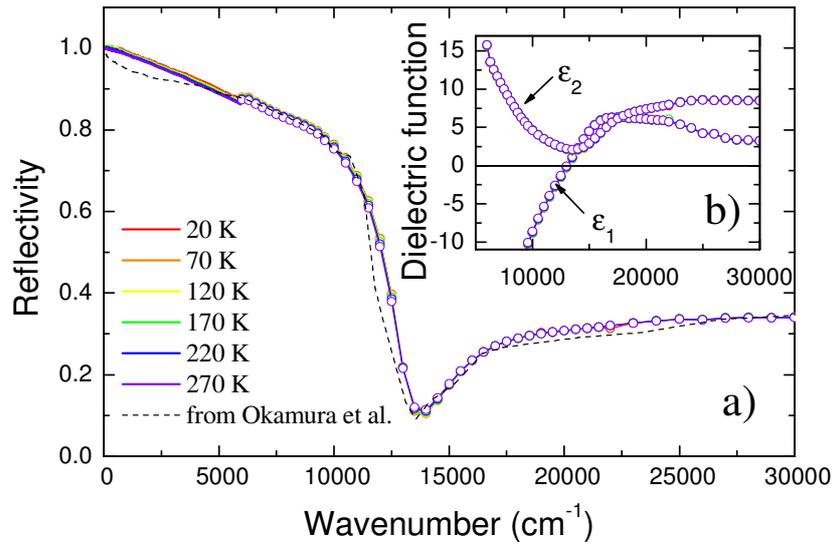

**Fig. 1 a) 1 b)**

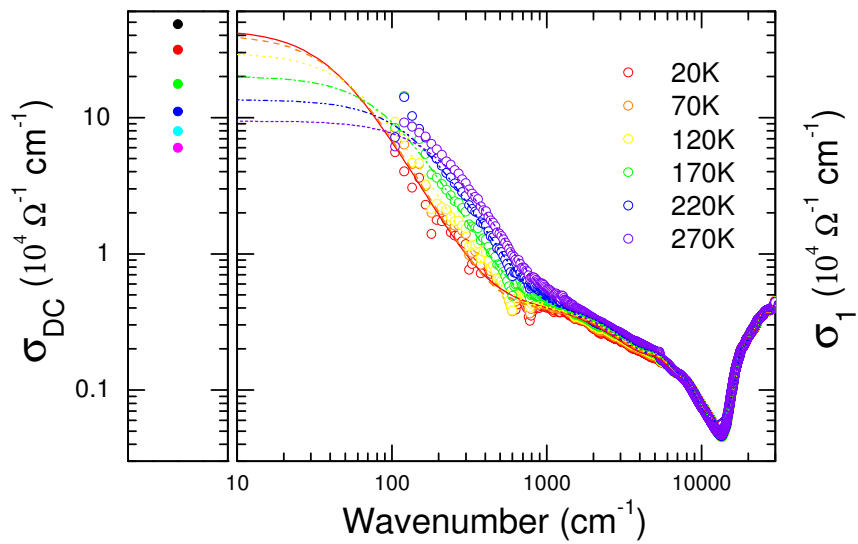

**Fig. 2.**

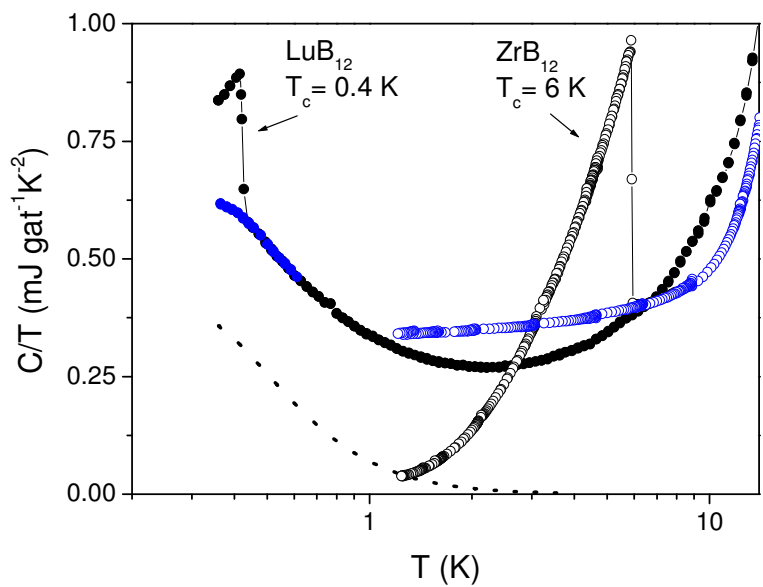

**FIG. 3.**

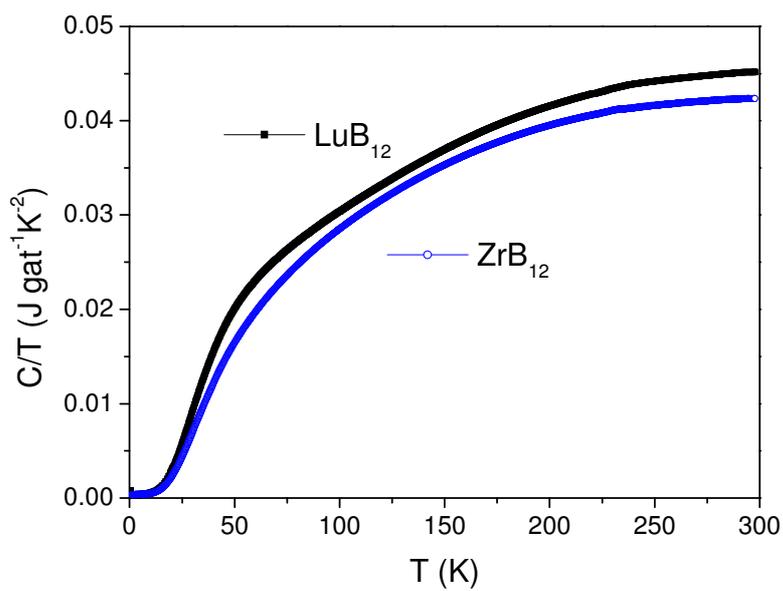

**FIG. 4.**

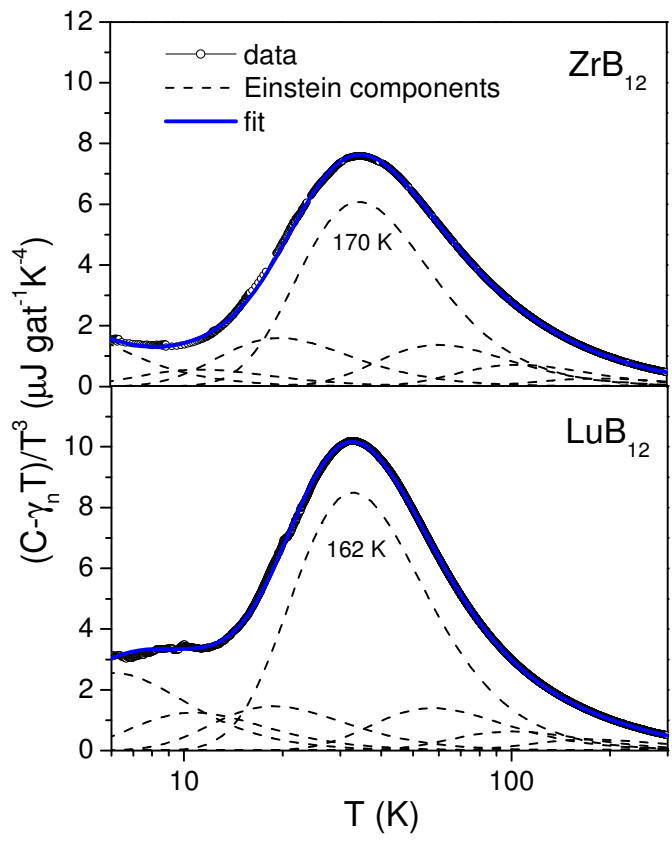

**FIG. 5.**

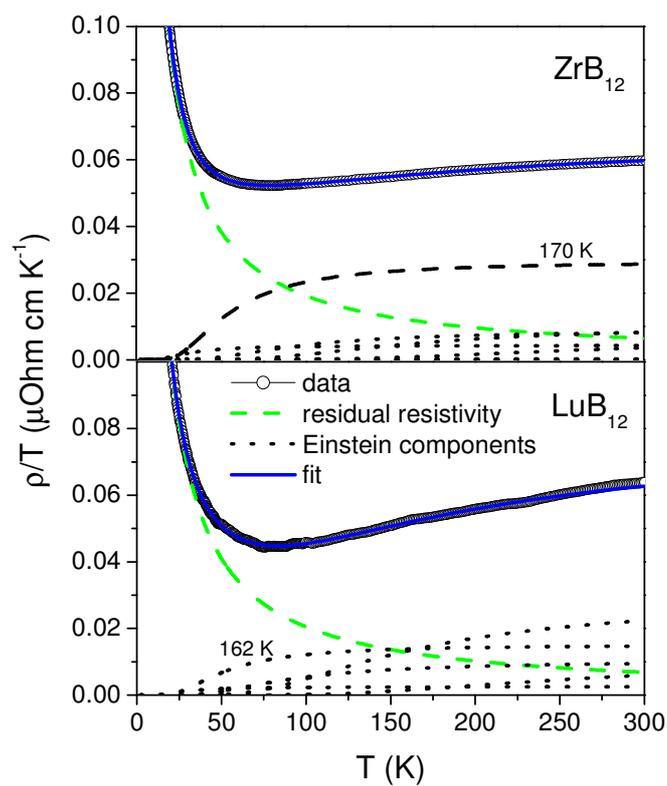

**FIG. 6.**

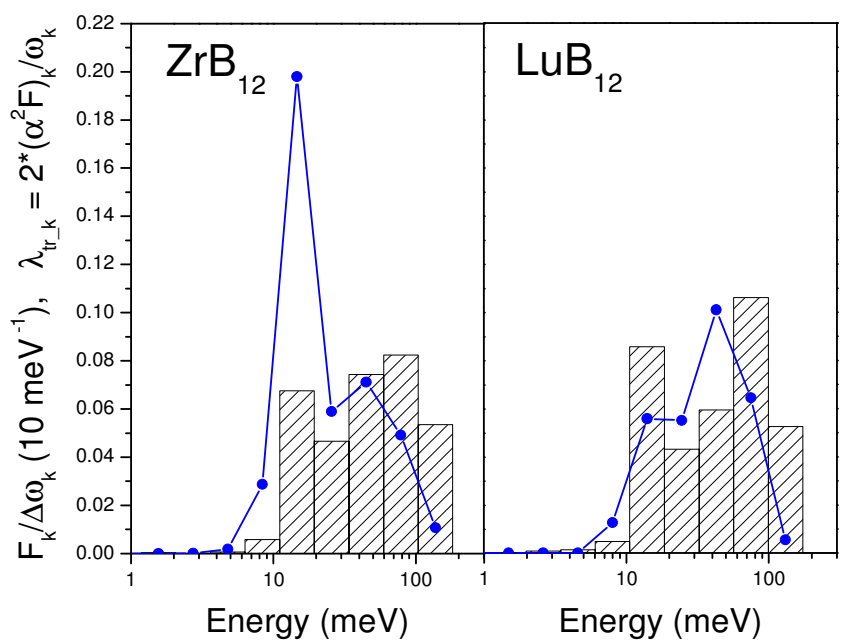

**FIG. 7.**

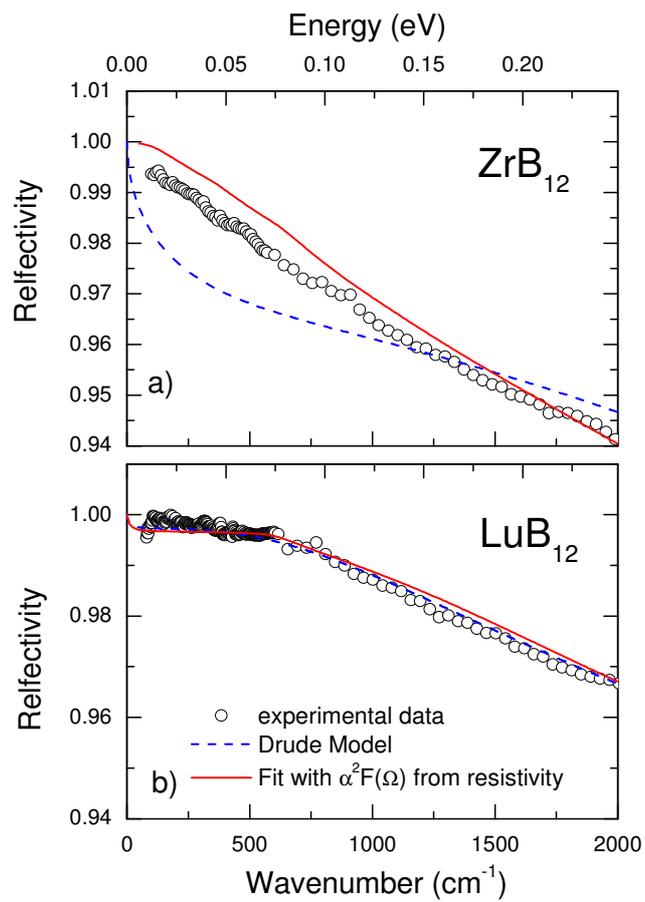

**Fig. 8.**